\documentclass[10pt]{article}
\usepackage{GECCO_proceedings2e}

\title{Numerical Replication of Computer Simulations: Some Pitfalls 
and How To Avoid Them
\thanks{\hspace{1ex}January 25, 2000 draft; submitted to GECCO 2000. Comments welcome!}
}

\author{
	Theodore C. Belding \\ 
	Center for the Study of Complex Systems \\
	University of Michigan \\ 
	Ann Arbor, MI 48109-1120 USA \\
	\texttt{Ted.Belding@umich.edu} \\ 
	\texttt{http://www-personal.umich.edu/\%7Estreak/}
}

\begin{document}

\maketitle


\begin{abstract} 
A computer simulation, such as a genetic algorithm, that uses IEEE
standard floating-point arithmetic may not produce exactly the same
results in two different runs, even if it is rerun on the same
computer with the same input and random number seeds. Researchers
should not simply assume that the results from one run replicate those
from another but should verify this by actually comparing the
data. However, researchers who are aware of this pitfall can reliably
replicate simulations, in practice. This paper discusses the problem
and suggests solutions.
\end{abstract}

\section{INTRODUCTION}

When we perform computer simulations, such as genetic algorithms, it
is often useful to be able to replicate a run exactly, so that those
results of the second run that we care about are exactly the same as
those of the first. This kind of replication is called {\em numerical
replication}~\cite{axtell:etal:1996}. For instance, if we notice a
strange result in a run, it is useful to be able to redo the run
exactly, using the same parameter settings and random number seeds,
but this time collecting additional data or perturbing the course of
the run in order to test hypotheses about what is causing the strange
results. Numerical replication can also be used to verify that
experimental results are not due to a bug or human error, and to
perform regression testing after making changes to a program.

However, a program that uses IEEE standard floating-point
arithmetic~\cite{ieee:1985,cody:etal:1984} may produce different
results on two different computers, even if the same input and random
number seeds are used.  In fact, there is no guarantee that it will
produce the same results when run twice on the same computer, or even
that a subexpression will have the same value when evaluated at two
different points during a single run. This is because the calculations
may be performed at different precisions each time, and the programmer
has little control over what precision is
used~\cite{priest:1997}. This can cause numerical replication to fail
unexpectedly. In the worst case, this can lead us to believe that two
different sets of results are the same, and thereby cause us to draw
incorrect conclusions.

Luckily, if we are aware of these pitfalls, we can reliably avoid them
in practice. We should not simply assume that two sets of data are the
same because we used the same input and random number seeds; instead,
we should always verify this empirically. Furthermore, we should
always record the computer platform and run-time and compile-time
parameters that we use along with the simulation data. This will make
numerical replication easier to achieve. We need tools to make both of
these tasks easy and automatic. Finally, we need to compile a
knowledgebase of heuristics for achieving numerical replication. In
the remainder of this paper, I discuss the problem further and justify
these recommendations.

\section{THE PROBLEM}
\label{sec:problem}

Computers perform arithmetic mainly on two kinds of numbers: integers
(such as 42) and real numbers (such as 3.14159).  There are various
possible ways to represent real numbers in a computer; almost all
modern computers use binary {\em floating-point}
representations~\cite{patterson:hennessy:1997,goldberg:1996,
goldberg:1991,knuth:1998}. This representation is essentially the same
as scientific notation, except in binary; besides representing real
numbers, it can also be used to represent very large
integers. Floating-point numbers are represented in the form $(-1)^{s}
\times 1.f \times 2^{x}$. The part $s$ is the {\em sign bit} ($0$ or $1$), 
$1.f$ is called the {\em significand} (an older term is {\em
mantissa}), $f$ is called the {\em fraction}, where $0 \leq f < 1$,
and $x$ is called the {\em exponent}. Almost all computer platforms
used today use IEEE 754 standard~\cite{ieee:1985,cody:etal:1984}
floating-point arithmetic. The only important exceptions are the Cray
X-MP, Y-MP, C90, and J90, the IBM /370 and 3090, and the DEC VAX; most
of these are disappearing rapidly~\cite{kahan:1996b}. This standard
has caused floating-point arithmetic to be much more reliable,
predictable, and portable.

\begin{figure}
\begin{verbatim}
#include <stdio.h>

int main() {
    double q;

    q = 3.0/7.0;
    if (q == 3.0/7.0) printf("Equal\n");
    else printf("Not Equal\n");
    return 0;
}
\end{verbatim}
\caption{Example program that may produce different results on different 
computer platforms, from Priest~\cite{priest:1997}}
\label{fig:example-program}
\end{figure}

\begin{figure}
\begin{verbatim}
#include <fpu_control.h>

void __attribute__ ((constructor))
enter_fpu_double_mode () {
    (void) __setfpucw ((_FPU_DEFAULT & 
        ~_FPU_EXTENDED) | _FPU_DOUBLE);
}
\end{verbatim}
\caption{Simply compile and link this C code with a program using the 
\texttt{gcc} compiler on x86 or m68k platforms to put the FPU in 
double-precision mode. Adapted from the g77 manual~\cite{g77:1999}. Other 
compilers should provide a similar way to do this. On x86, more needs to 
be done to completely emulate double precision; see the text for details.}
\label{fig:set-fpu-double-mode}
\end{figure}

However, the standard does not guarantee that a program will produce
the same results when run on two different
computers~\cite{priest:1997}.  This is because different computers may
perform floating point calculations differently, even if the computers
all follow the IEEE standard.  The standard specifies three different
precisions for floating-point arithmetic: {\em single precision} (32
bits long), {\em double precision} (64 bits) and {\em double extended
precision} (also called simply {\em extended precision}, 80 or more
bits).  Different computer platforms support these precisions to
different extents. 

We may get different results, for instance, when we run a simulation
once on a Hewlett-Packard (HP) PA-RISC workstation running HPUX and once
on an Intel x86 PC running Linux or MS Windows. The HP workstation
uses single-precision and double-precision floating point arithmetic,
while the x86 uses IEEE 80-bit extended-precision floating-point
arithmetic by default. (The Motorola 680x0 (m68k) is another CPU
family that uses 80-bit extended precision; it was used in the first
Macintosh computers.)  Figure~\ref{fig:example-program} shows an
example program that may produce a different result on each platform,
depending on the compiler and the compile-time settings. An HP
workstation will print ``Equal'', while an x86 computer may print
either ``Equal'' or ``Not Equal'', depending on the compiler and
compile-time options that are used~\cite{priest:1997}.

If the results of a simulation depend on many floating-point
calculations, this difference in precision may cause the two runs to
produce wildly different results. This is particularly likely in
simulations of complex systems, such as a genetic algorithm, where the
simulation's precise trajectory is highly sensitive to the initial
conditions and to the stream of random numbers. Even if the different
runs produce the same qualitative results, the numeric results may
differ.

This may occur with any program that uses native IEEE floating-point
arithmetic, written in any language, on any computer or operating
system.  Discrepancies may also occur in integer arithmetic, but
only if a program makes unwarranted assumptions about the size or
representation of integer variables (for example, assuming that C
variables of type \texttt{int} are 32 bits long).

Both the x86's and the m68k's floating-point unit (FPU) can be
switched into ``single-precision'' or ``double-precision'' mode (see
Figure~\ref{fig:set-fpu-double-mode}). This solves this particular
problem on the m68k.  Unfortunately, even when the x86 is in one of
these modes, it will still produce different results than an HP or
similar workstation would, since its internal registers will still use
more bits of precision for the exponents (15 bits instead of 8 or
11)~\cite{java:1998}. To reduce the exponent range to be the same as
that in ``pure'' single or double precision, the result must be stored
to memory from the x86 FPU's internal registers, and then reloaded
from memory into the FPU. This will cause the computation to be two to
four times slower than native floating-point arithmetic. If the
\texttt{gcc} compiler is being used, this can be accomplished by using
the \texttt{-ffloat-store} compiler option. However, there may still
be a discrepancy on the x86 in the last bit of about $10^{-324}$
because of double rounding, if the floating-point operation is a
multiplication or division. To avoid this, one of the operands must be
scaled down before the operation by
$2^{x_{\mathrm{max}_{\mathrm{extended}}} -
x_{\mathrm{max}_{\mathrm{double}}}}$, where
$x_{\mathrm{max}_{\mathrm{extended}}}$ is the maximum possible
exponent for extended precision and
$x_{\mathrm{max}_{\mathrm{double}}}$ is the maximum possible exponent
for double precision, and the result must be scaled back up by the
same amount afterwards. This additional scaling adds only marginally
to the computation time~\cite{java:1998}.
 
By using this technique, replication problems can be made much less
likely, at the expense of computation speed.  However, it will not
guarantee that such problems will not occur. In fact, a program may
produce different results when run twice on the same computer, even if
the same input and random number seeds are used.  This is because the
results produced by a program depend not only on the computer's
floating-point unit and operating system but also on the compiler, the
compile-time options, the compile-time and run-time libraries
installed, and the input (here I include the date, the run-time
environment, and the random number seeds). For instance, the
discrepancy may occur if we run a simulation twice on a x86 computer,
where the simulation is compiled the first time to store
floating-point results to memory, and the second time to keep the
results in the FPU's internal registers. Also, the libraries of
mathematical functions such as $\log$ and $\sin$ may produce different
results on different platforms~\cite{kahan:1996b} and may also differ
from version to version on the same platform. (The IEEE standard only
contains specifications for the square root function $\sqrt{x}$.) The
IEEE standard also does not completely specify the accuracy conversion
between binary and decimal representations. It is even technically
possible that the results may depend on what other programs are
running on the computer, or on bugs in the program, compiler, or
libraries --- this is especially true if the program is not carefully
designed and implemented. Therefore, each time a simulation is run, it
is prudent to act as if it were run on a different computer, even if
the computer is in fact always exactly the same.

A related issue is that if a floating-point expression occurs more
than once in different locations in a program, it may be evaluated to
different precisions each time it is used during a single
run~\cite{priest:1997}. For example, on an x86 computer the compiler
may choose whether to keep a result in extended precision in the FPU
or store it in double precision to memory based on the optimization
level, the number of free floating-point registers, whether the result
will be used as the argument to a function, and many other
factors. (The forthcoming C99 ANSI/ISO C standard will guarantee that
if the expression is stored in a variable, the same precision will be
used whenever the variable's value is evaluated.)  Besides
complicating numerical replication, this may cause problems if the
program assumes that the expression always evaluates to the same value
during the course of a run. The \texttt{fcmp}
package~\cite{belding:2000b} implements Knuth's~\cite{knuth:1998}
suggestions for safer floating-point comparisons, which can be used to
avoid this.

Finally, some CPUs, such as the PowerPC, provide an operation called
{\em fused multiply-add} that can perform the operation $\pm a x \pm
b$ in a single instruction. If this instruction is used, a different
result may be produced than if it is not used, since there is one less
rounding step~\cite{priest:1997}. Also, in expressions such as $\pm a x
\pm b y$ it is ambiguous which side is evaluated first (and hence
rounded). Therefore, this instruction must not be used in certain
algorithms, for instance when multiplying a complex number by its
conjugate~\cite{kahan:1996a,kahan:1996b}. Unfortunately, many
compilers make it difficult for the programmer to specify whether this
instruction should be allowed or inhibited in a program.

Guaranteeing that two runs of a program will produce exactly the same
results is extremely difficult and may be impossible in practice.
Every component which might affect the results would have to be
guaranteed to be the same for both runs; none of these components
could ever be changed or upgraded unless the new version could be
shown to have no effect on the results.  On the one hand, determining
the version of every component on a computer and recording all of this
information with the simulation data would be extremely expensive in
time and storage.  On the other hand, it will be extremely difficult
to weed out false positive results when testing whether two computers
have different components: The fact that one of two otherwise
identical computers has a copy of the game Quake installed probably
will not affect whether a simulation will produce identical results on
the two machines, but it will be difficult to prove this. Finally, the
date is always changing, and this might have unforeseen effects on a
program's behavior. (Consider the recent Y2k problem, or the bug that
depended on the phase of the moon~\cite{raymond:1993}.)

\section{RECOMMENDATIONS}

If guaranteeing that we can numerically replicate a run is not an option,
what can we do? I suggest that instead of asking how we can guarantee
replication, we should ask two different questions: First, what is the
worst-case result that can occur because of this problem, and how can
we avoid it? Secondly, how can we make numerical replication easier to
achieve and more reliable?

The worst thing that can happen when we try to numerically replicate a
run is that we mistakenly believe that the replicated results are
exactly the same as the original, when they are in fact different. Our
main concern should be to avoid this mistake. Luckily, there is an
easy way to avoid it: Simply compare the data sets. If they are
empirically identical, we are done. (Of course, if we do not record
enough data from each run, it is possible that the runs' actual
trajectories may be different, even though the data are the same.) 

Therefore, we need a set of easy-to-use tools to compare results from
two runs, and we should use these tools even if the runs were done on
the same computer, as a sanity check. In some cases, where entire
files need to match exactly, a utility such as the Unix \texttt{diff}
command may suffice. In other cases, I suggest using
Rivest's~\cite{rivest:1992} MD5 message digest algorithm.  This
algorithm produces a short string (called a {\em hash}) that is easy
to store with the data that it is computed from.  Instead of comparing
entire files, only the hash string from each file needs to be
compared. If the data files clearly mark comments and other data that
we do not need to replicate, such as the date of the run, then it is
easy to write a short Perl program~\cite{wall:etal:1996,winton:1996} to
compute an MD5 hash string from a data file, ignoring such extraneous
information. (One common convention for marking comments in text files
is to put a pound sign `\texttt{\#}' at the beginning of each comment
line.)  If it is necessary to ensure that a dataset has not been
tampered with in any way, there are cryptographically secure methods,
such as signing the data set with
PGP~\cite{zimmermann:1995,garfinkel:1994} or GnuPG~\cite{koch:1999},
or using another message digest algorithm, such as
RIPEMD-160~\cite{dobbertin:etal:1996,bosselaers:1999}. (MD5 should
{\em not} be used for this purpose~\cite{robshaw:1996}!)

Often, using the same input and random number seed will be all that is
necessary to numerically replicate a run.  Sometimes, however, this
will not suffice. In this case, we can almost always replicate the
results by tweaking a few special compile-time or run-time parameters
(such as what precision the FPU uses).  Experience suggests that
numerical replication is usually easy to achieve in practice, even
though it may be impossible to guarantee.  In some cases, it may be
necessary to rerun the simulation on the same computer platform that
was used originally.  For instance, if a simulation is run on an x86
platform using extended precision, it will be difficult to numerically
replicate the results on any platform other than an x86 or m68k using
extended precision.  In addition to the techniques for comparing
results mentioned previously, we need a set of heuristics for
numerical replication, such as a list of compile-time and run-time
parameters that often need to be tweaked.  One such heuristic is the
technique for emulating double-precision floating-point on x86
computers described in Section~\ref{sec:problem}.

To make numerical replication easier, the compile-time and run-time
parameters that were used should be stored with a simulation's results,
along with information such as the program and compiler versions, the
date, the name of the machine being used, the platform and operating
system, etc. In addition to tools for comparing simulation results, we
also need tools that make storing this kind of information easy and
automatic. (Perl~\cite{wall:etal:1996} is an example of a program that
stores a great deal of configuration information at compile-time; the
information is accessible under Unix by running
\texttt{perl -V}.) A researcher can then use this information when
trying to numerically replicate the run. For example, if a simulation
is run on an x86 computer using extended precision, it is important to
record this fact.

A future release of Drone~\cite{belding:2000a} will include tools for
recording and comparing MD5 hashes of data files and for recording
compile-time and run-time parameters of simulation programs. I hope
that making these tools available will encourage researchers to use
them every time they run a simulation.

\section{CONCLUSION}

In summary, these are problems that everyone doing computer
simulations should be aware of, but they are not insurmountable. In
practice, a few simple techniques should be sufficient to avoid
problems. First, we should never assume that the results from two
simulation runs are identical because they used the same parameters
and random number seed, even if they are run on the same computer. We
should always verify this, either by comparing the relevant results
directly or by comparing the MD5 hash strings of the two
datasets. This verification process should be made so convenient that
there is no reason not to do it. Secondly, we should compile a
knowledgebase of likely parameters that can be tweaked to achieve
numerical replication, if simply redoing a run with the same input and
random seeds does not suffice. Finally, we should always store the
compile-time and run-time parameters that we use. We need tools to
make this convenient and automatic, as well.

\section{FURTHER READING}

For a technical discussion of this problem, see
Priest~\cite{priest:1997} and the Java Grande Forum Numerics Working
Group's draft report~\cite{java:1998}. For a gentle introduction to
floating-point arithmetic in general, see Patterson and
Hennessy~\cite{patterson:hennessy:1997} or
Goldberg~\cite{goldberg:1996}; for a more technical discussion, see
Goldberg~\cite{goldberg:1991} or Knuth~\cite{knuth:1998}. The IEEE 754
floating-point standard is published in~\cite{ieee:1985}; for a
readable account see Cody, et al.~\cite{cody:etal:1984}. Cody and
Coonen~\cite{cody:coonen:1993} give C algorithms to support features
of the standard.  Kahan and Darcy~\cite{kahan:darcy:1998} and
Darcy~\cite{darcy:1998} argue that it is undesirable to enforce exact
replicability across all computing platforms, and
Kahan~\cite{kahan:1998} gives an example of differences in
floating-point arithmetic in different versions of \texttt{Matlab} on
various platforms.  Axtell, et al.~\cite{axtell:etal:1996} discuss the
differing degrees to which a simulation can be replicated.  See
Rivest~\cite{rivest:1992} and Robshaw~\cite{robshaw:1996} for
information on the MD5 message digest algorithm; for a Perl interface
to MD5, see Winton~\cite{winton:1996}. For information on RIPEMD-160, a
more secure replacement for MD5, see Bosselaers~\cite{bosselaers:1999}
or Dobbertin~\cite{dobbertin:etal:1996}.

\section{ACKNOWLEDGMENTS}

I am grateful to the members of the \texttt{egcs} and \texttt{gcc}
mailing lists for answering questions and providing information and
references. I also thank the other members of the University of
Michigan Royal Road group (John Holland, Rick Riolo, Bob Lindsay,
Leeann Fu, Tom Bersano-Begey, and Chien-Feng Huang) for their
comments and encouragement.

\bibliographystyle{plain} 
\bibliography{numerical-replication} 
\end{document}